# Orthogonal Chirp Division Multiplexing

Xing Ouyang and Jian Zhao

*Abstract*—Chirp waveform plays a significant role in radar and communication systems for its ability of pulse compression and spread spectrum. This paper presents a principle of orthogonally multiplexing a bank of linear chirp waveforms within the same bandwidth. The amplitude and phase of the chirps are modulated for information communication. As Fourier transform is the basis for orthogonal frequency division multiplexing (OFDM), Fresnel transform underlies the proposed orthogonal chirp division multiplexing (OCDM). Digital implementation of the OCDM system using discrete Fresnel transform is proposed. Based on the convolution theorem of the Fresnel transform, the transmission of the OCDM signal is analyzed under the linear time-invariant or quasi-static channel with additive noise, which can generalize typical linear transmission channels. Based on the eigen-decomposition of Fresnel transform, efficient digital signal processing algorithm is proposed for compensating channel dispersion by linear single- tap equalizers. The implementation details of the OCDM system is discussed with emphasis on its compatibility to the OFDM system. Finally, simulation are provided to validate the feasibility of the proposed OCDM under wireless channels. It is shown that the OCDM system is able to utilize the multipath diversity and outperforms the OFDM system under the multipath fading channels.

This work was supported by the Science Foundation Ireland under Grants 11/SIRG/I2124.

Xing Ouyang and Jian Zhao are with Tyndall National Institute, Lee Maltings Complex, Dyke Parade, Cork, Ireland. (e-mail: xing.ouyang@tyndall.ie, jian.zhao@tyndall.ie).

Xing Ouyang is also with the School of Electrical and Electronic Engineering, University College Cork, 60 College Rd, Cork, Ireland.

## I. INTRODUCTION

CHIRP relates to a signal whose phase evolves along with time in a certain manner, and it can be found almost everywhere. For example, the spatial frequency of the near-field Fresnel diffraction pattern increases linearly with the distance to the origin of the screen, in Fig. 1 (a) and (b) [1-4]. In radar systems, the chirp signal is frequency modulated and swept over a wide spectrum with a constant amplitude. By correlating the echoes bouncing from target, pulse compression is achieved as if 'short' pulses are emitted and received. The position of the target can be resolved from the temporally delayed pulses [5-7]. In spread-spectrum systems, digital binary information is encoded by modulating the frequency of a carrier, sweeping the frequency linearly from low to high as 'up-chirp' for bit '1' or from high to low as 'down-chirp' for bit '0', as shown in Fig. 1 (c) and (d) [8-13]. The chirp signal uses a wideband spectrum for transmission and is resistant to detrimental effects, such as channel noise, multipath fading and Doppler effects within the mobile radio channel.

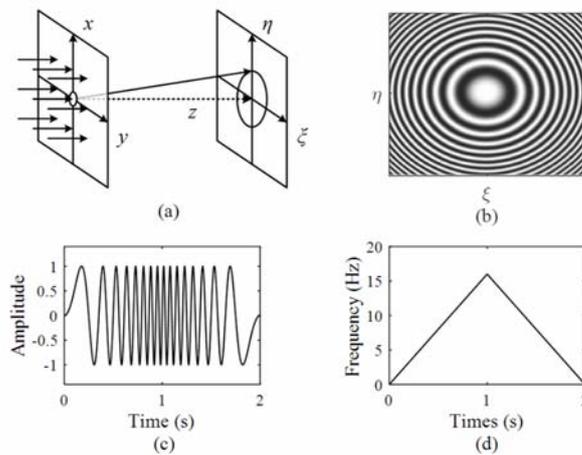

Fig. 1. Illustration of chirps: (a) Near-field Kirchhoff-Fresnel diffraction of a circular aperture, and (b) its diffraction pattern on the second plate; (c) linear chirp waveform which is 'up-chirp' from 0 to 1 second and is 'down-chirp' from 1 to 2 second, and (d) its spectrogram.

For the purpose of communication, the chirp signal is of spread-spectrum approach which guarantees secure and robust communication for applications including military, underwater and aerospace communications [14-16]. In the recent ultra-wideband (UWB) standard by the IEEE 802.15.4a group [17], the chirp signal is adopted to satisfy the requirement of FCC on the radiation power spectral mask for the unlicensed UWB systems. Meanwhile, the chirp signal achieves what a UWB system is supposed to do, i.e., ranging, measuring and communicating [18-20].

Conventionally, the chirp signal is usually generated by analog devices in filter approach by surface acoustic wave (SAW) devices [21-25] or in frequency-modulation approach by voltage controlled oscillator taking the advantages of the mature CMOS technology [26-28]. In the chirp spread-spectrum (CSS) or frequency-modulated systems, a broad spectrum is occupied for modulating information. Spectral efficiency is sacrificed for higher processing gain, the capability of multipath resolution, and other merits of the chirp signal as well. In a given period $T$ and bandwidth $B$, a single chirp is modulated for transmission as shown in Fig. 2 (a). If there exists more than one chirp within the same period and bandwidth, inter-chirp interference occurs. As a result, the chirp or frequency-modulated signal is attractive for low-rate applications in which reliable communication is in priority.

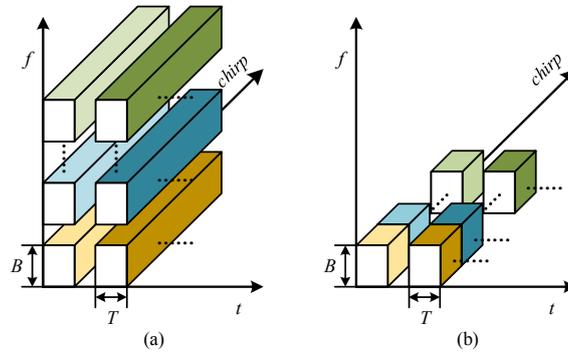

Fig. 2. Illustration of (a) the multi-code chirp waveform and (b) the orthogonal chirp division multiplexing in the temporal-frequency-chirp dimension.

To increase the data rate of the UWB system, multi-code UWB is proposed by dividing the entire spectrum into a bank of spectrally separated chirps [18], see Fig. 2 (a). Each chirp is modulated in binary codes, such as, the Walsh-Hadamard code. At the receiver, the transmitted information is recovered by retrieving the orthogonal codes from the modulated chirps over the entire bandwidth.

Heuristically, in this paper we came to think about if there is a way of synthesizing a bank of chirps in the same period and bandwidth without any interference, and both of the amplitude and phase of the chirp can be used for modulation. If yes, so can we implement such system by using the advanced digital signal processing (DSP) technology in the digital domain rather than in the analog domain?

In this paper, we introduce a principle of orthogonally multiplexing a number of chirp waveforms within the same bandwidth to achieve the maximum communication rate, termed as orthogonal chirp division multiplexing (OCDM). As shown in Fig. 2 (b), a bunch of chirp waveforms overlap temporally and spectrally in the OCDM. The phase and/or amplitude of each chirp signal are modulated, for example, by phase shift keying (PSK) or quadrature amplitude modulation (QAM). The modulated chirps are orthogonal along the chirp dimension, without interference to each other for transmission.

We will show that the fundamental mechanism behind the OCDM is the Fresnel transform, just as the Fourier transform in orthogonal frequency division multiplexing (OFDM). In this paper, digital implementation of the OCDM system using discrete Fresnel transform (DFnT) is presented. Specifically, the inverse DFnT (IDFnT) generates the OCDM signal at the transmitter, and the DFnT recovers the OCDM signal at the receiver.

We then analyze the transmission of the OCDM signal under linear time-invariant (LTI) or quasi-static channels which can generalize most of the practical linear communication channels, like coaxial cables, optical fibers and mobile radio channels etc. Based on the convolution property of Fresnel transform, it is shown that the channel distortion can be compensated by using either time-domain or frequency-domain equalizer.

Moreover, according to the eigen-decomposition of DFnT, a more efficient channel equalization algorithm is proposed for the OCDM system. The algorithm is based on single-tap frequency domain equalization (FDE), and it is more efficient than the aforementioned two equalization schemes.

This paper is organized as follows. In Section II, the Fresnel transform and its discrete form,

DFnT are introduced for supporting the OCDM. In Section III, mathematical description of the proposed OCDM system is presented, and its transmission under LTI channels is investigated analytically. The efficient linear single-tap FDE is presented in Section IV. Discussions in terms of implementing the OCDM system and its compatibility to the OFDM system are provided in Section V. In Section VI simulations are performed to evaluate the performance of the OCDM system under wireless multipath fading channels, and Section VII finally concludes this paper.

In this paper, we consider the system in baseband representation [Chapter 2, 29]. Some basic mathematical notations are given here. The imaginary unit is $j = \sqrt{-1}$ and the operators $*$ and $\circledast$ denote linear and circular convolution, respectively. The discrete system model is usually in matrix form. Matrices are upright bold in uppercase while vectors are italic bold in lowercase. The ($m$, $n$)-th entry of matrix $\mathbf{H}$ and the $n$-th entry of vector $\boldsymbol{h}$ are H($m$, $n$) and $h(n)$. The superscripts $(\cdot)^*$, $(\cdot)^T$ and $(\cdot)^H$ denote complex conjugate, transpose and Hermitian transpose operators, respectively. The Dirac delta function is $\delta(t)$ and the Kronecker delta is $\delta(n)$, depending on the variable being continuous or discrete.

## II. FRESNEL TRANSFORM

In this section, we will introduce the Fresnel transform and the discrete Fresnel transform as the basis of the OCDM system in the following sections. The linear (circular) convolution property of the Fresnel transform (DFnT) is also presented to support the mathematical model of the OCDM signal transmission under LTI channels.

### A. Fresnel Transform

Fresnel transform is an integral transformation originating from classical optics [1-4]. It is the formula that mathematically describes the behavior of the near-field optical diffraction. As shown in Fig. 1 (a), when a monochromatic plain wave with wavelength $\lambda$ encounters a slit (grating) which is comparable in size to $\lambda$, the resulting near-field diffraction pattern observed on a plate at

distance $z$ is given by

$$\hat{s}(\tau) = \mathcal{F}_a\{s(t)\}(\tau) = \frac{e^{-j\frac{\pi}{4}}}{\sqrt{a}} \int s(t) e^{j\frac{\pi}{a}(\tau-t)^2} \mathrm{d}t, \qquad (1)$$

where $\mathcal{F}_a\{\}(\cdot)$ denotes the Fresnel transform, and $a = \lambda z$ is the normalized Talbot distance, $s(t)$ is the complex transmittance of the grating. In Equ. (1), $\hat{s}(t)$ is called the Fresnel transform of $s(t)$. Equivalently, the Fresnel transform can be expressed in the form of convolution [30] as

$$\hat{s}(\tau) = s(\tau) * \varphi_a(\tau), \qquad (2)$$

where

$$\varphi_a(t) = e^{-j\frac{\pi}{4}} e^{j\frac{\pi}{a}t^2}. \qquad (3)$$

Fresnel transform, as well as the Fourier transform and fractional Fourier transform, is the special case of linear canonical transform (LCT) [31-34]. Fresnel diffraction is a fundamental phenomenon in optics and quantum physics [35-37], while it is studied almost exclusively by using Fourier analysis. In 1994, Gori visited the Fresnel transform in his chapter "Why Fresnel is so little known?" from the perspective of the Fresnel transform itself rather than from the Fourier analysis approach [38]. And the important but almost 'neglected' property [Theory (1), 38], the Fresnel transform of a linear convolution is

$$\begin{aligned} \hat{r}(\tau) &= \mathcal{F}_a\{h(t) * s(t)\}(\tau) \\ &= \hat{h}(\tau) * s(\tau) = h(\tau) * \hat{s}(\tau). \end{aligned} \qquad (4)$$

In Equ. (4), it states that the Fresnel transform of a convolution equals either one convolving with the Fresnel transform of the other. This is different from the convolution theorem of the Fourier transform which says that the Fourier transform of a convolution equals the product of the Fourier transforms.

## B. Discrete Fresnel Transform

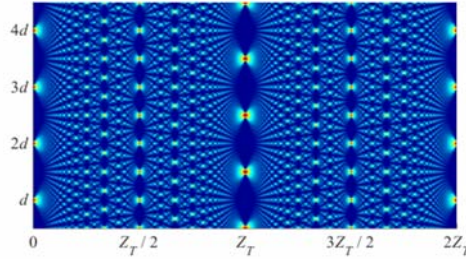

Fig. 3. Illustration of Talbot effects; the DFnT matrix of size $N$ represents the optical field of the light spots at the fraction of Talbot distance $Z_T/N$.

The discrete form of Fresnel transform, DFnT relates to the Talbot effect which is the periodic grating of the Fresnel diffraction [1-4, 37, 39-42], as shown in Fig. 3. The DFnT matrix gives the optical field coefficients of the Talbot image or called self-image at the fraction of Talbot distance, $z = Z_T/N$, where

$$Z_T = d^2/\lambda \tag{5}$$

is the Talbot distance and $d$ is the distance of repeated gratings.

In previous works, the DFnT is formulated for describing the coefficients of Talbot image. There exists degeneracy in the DFnT matrices: the size of DFnT matrices is $N/2$ if $N \equiv 0$ and 2 (mod 4), while the size of DFnT matrices is $N$ if $N \equiv 1$ and 3 (mod 4) [39, 41, 43-45]. This degeneracy hinders the application of DFnT as a general mathematical tool.

In the recent work, the DFnT is derived in [46] without such degeneracy. The $(m, n)$-th entry of the $N$ by $N$ DFnT matrix $\mathbf{\Phi}$ is defined as

$$\Phi(m,n) = \frac{1}{\sqrt{N}} e^{-j\frac{\pi}{4}} \times \begin{cases} e^{j\frac{\pi}{N}(m-n)^2} & N \equiv 0 \,(\mathrm{mod}\,2) \\ e^{j\frac{\pi}{N}\left(m+\frac{1}{2}-n\right)^2} & N \equiv 1 \,(\mathrm{mod}\,2) \end{cases}. \tag{6}$$

The DFnT matrix is unitary, and its other important properties, such as its eigen-decomposition, can be found in [46]. It should be noted that the representations of the DFnT is slightly different for the DFnT of even and odd $N$.

The DFnT possesses the circular convolution property which says that the DFnT of the circular convolution of two sequences equals either one convolving with the DFnT of the other. Given two length-$N$ vectors $\boldsymbol{h}$ and $\boldsymbol{s}$ and two $N$ by $N$ circulant matrices $\mathbf{H}$ and $\mathbf{S}$ whose first columns are $\boldsymbol{h} = [h(0), h(1), \ldots, h(N-1)]^T$ and $\boldsymbol{s} = [s(0), s(1), \ldots, s(N-1)]^T$, the circular convolution $r(n) = h(n) \circledast s(n)$ in matrix form is

$$\boldsymbol{r} = \mathbf{H}\boldsymbol{s} = \mathbf{S}\boldsymbol{h}. \tag{7}$$

The DFnT of the circular convolution is

$$\hat{\boldsymbol{r}} = \boldsymbol{\Phi}\boldsymbol{r} = \mathbf{H}\hat{\boldsymbol{s}} = \mathbf{S}\hat{\boldsymbol{h}}. \tag{8}$$

where $\hat{\boldsymbol{s}}$ and $\hat{\boldsymbol{h}}$ are the DFnTs of $\boldsymbol{s}$ and $\boldsymbol{h}$, respectively. It can be observed that the DFnT of circular convolution in Equ. (8) is the discrete analogy of Equ. (4).

## III. Orthogonal Chirp Division Multiplexing

In this section, after a brief review of CSS systems, we will show that the Fresnel transform provides a theoretical framework for OCDM, and we also proposed the digital implementation of the OCDM system.

### A. Chirp Spread Spectrum

Most applications consider the frequency modulated (chirp) signal whose frequency evolves linearly or equivalently whose phase quadratically over time,

$$\psi(t) = e^{j\left(\pi \alpha t^2 + \varphi_0\right)}, \tag{9}$$

where $\alpha$ is the chirp rate and $\varphi_0$ is an initial phase. Its instantaneous frequency is

$$f(t) = \frac{1}{2\pi} \frac{\mathrm{d}}{\mathrm{d}t}\left[\pi \alpha t^2 + \varphi_0\right] = \alpha t. \tag{10}$$

If the chirp signal is temporally limited within some period $T$, the bandwidth of the chirp signal $B$ is determined by the chirp rate $\alpha$ and the period $T$, i.e.,

$$B \propto (\alpha / T). \tag{11}$$

The time-bandwidth product $\alpha \propto B \times T$ indicates the processing gain of a chirp signal.

In the CSS system, the time-bandwidth product $BT \gg 1$, and thus $B \gg R_s$, where $R_s = 1/T$ is the symbol rate. It means that the larger the processing gain is, the less the spectral efficiency becomes. Moreover, in the conventional CSS system, the chirp signal is modulated by analog devices. The advanced modulation formats like QAM cannot be implemented to enhance the spectral efficiency. In the following subsection, we will present the principle of OCDM to maximize the spectral efficiency of CSS system.

## B. Principle of Orthogonal Chirp Division Multiplexing

To introduce the Fresnel transform for OCDM, some constraints should be raised. Firstly, the chirp waveforms used for information transmission should be time-limited. Secondly, we need to adapt the spatial Talbot effect from the optics into the temporal counterpart for OCDM. Based on Equ. (5), we define the temporal Talbot distance $Z_T$ as

$$Z_T = \frac{T^2}{\lambda} \tag{12}$$

where $T$ is the period of a chirp waveform. It can be observed that the temporal parameter $T$ replaces the periodic distance $d$ in Equ. (5). Supposing that there are $N$ chirp waveforms, one can obtain the chirp waveform by substituting the fraction of the Talbot distance $z = Z_T/N$ into Equ. (3), and thus $a = T^2/N$. The 'root' chirp waveform is defined as

$$\psi_0(t) = \Pi_T(t)\varphi_a^*(t)\Big|_{a=\frac{T^2}{N}} \tag{13}$$
$$= e^{j\frac{\pi}{4}}e^{-j\pi\frac{N}{T^2}t^2}, \quad 0 \le t < T$$

where

$$\Pi_T(t) = \begin{cases} 1 & 0 \le t < T \\ 0 & \text{otherwise} \end{cases}, \tag{14}$$

is the rectangular function.

It can be seen that Equ. (13) is the chirp signal with chirp rate $\alpha = N/T^2$, and its time-bandwidth product is about $BT = N$. One can get a set of $N$ chirp waveforms by using the root chirp in Equ. (13), and the $k$-th, $k = 0, 1, …, N − 1$, chirp waveform is

$$\psi_k(t) = \Pi_T(t)\varphi^*_{T^2/N}\left(t - k\frac{T}{N}\right)$$
$$= e^{j\frac{\pi}{4}}e^{-j\pi\frac{N}{T^2}\left(t - k\frac{T}{N}\right)^2}, \quad 0 \leq t < T. \tag{15}$$

It can be readily proved that the chirp waveforms $\psi_k(t)$ in Equ. (15) are mutually orthogonal,

$$\int \psi^*_m(t)\psi_n(t)\mathrm{d}t$$
$$= \int_0^T e^{j\pi\frac{N}{T^2}\left(t - m\frac{T}{N}\right)^2} e^{-j\pi\frac{N}{T^2}\left(t - n\frac{T}{N}\right)^2} = \delta(m - n). \tag{16}$$

Equ. (15) formulates a set of $N$ orthogonal chirp waveforms in a given bandwidth and period. Thus, the spectral efficiency of the OCDM increases by $N$ over the CSS system. To give an illustrative comparison between the OFDM and the OCDM, Fig. 4 presents the linear exponential waveforms in OFDM that are mutually orthogonal in frequency [47-50], and the quadratic exponential waveforms in OCDM that are mutually orthogonal in the dimension of chirp.

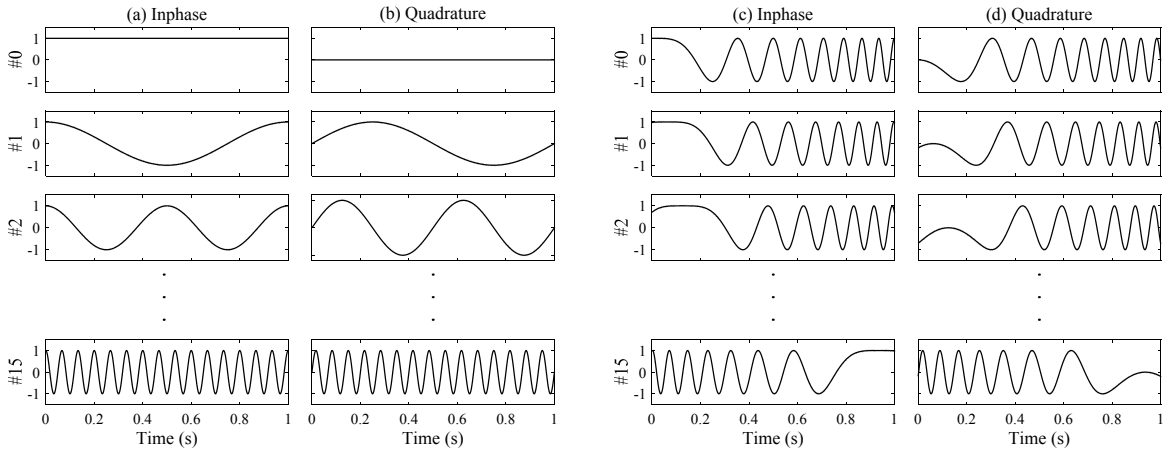

Fig. 4. Illustration of the families of 16 orthogonal waveforms in OFDM and OCDM. (a) The inphase and (b) the quadrature components of the orthogonal linear exponential waveforms in OFDM, and (c) the inphase and (d) the quadrature components of the orthogonal chirp waveforms in OCDM.

In the OCDM system, the amplitude and phase of each chirp can be used for modulation. Thus, pulse amplitude modulation (PAM), PSK, and QAM can be employed. Depending on the modu-lation formats, symbols are chosen from a codebook $\chi$ to encode the information bits. Similar to the OFDM symbol that consists of a bank of subcarriers transmitted block by block, the modulated chirps are transmitted in block as well. In an OCDM block, the $k$-th symbol modulating the $k$-th chirp is $x(k) \in \chi$. A bank of synthesized and modulated chirp signal, see Fig. 5 (a), is thus

$$s(t) = \sum_{k=0}^{N-1} x(k)\psi_k(t) \quad 0 \leq t < T. \tag{17}$$

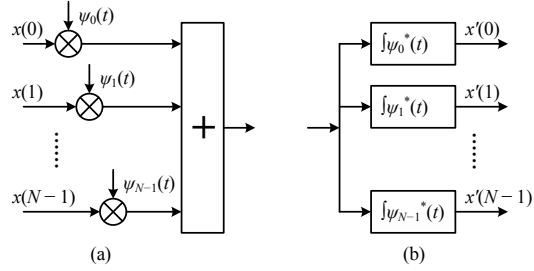

(a)           (b)

Fig. 5. Schematic diagram of the OCDM transceiver (a) multiplexing and (b) demultiplexing a bank of $N$ modulated orthogonal chirp waveforms.

According to Equ. (16), $x(m)$ can be extracted by the matched filter to the $m$-th chirp as shown in Fig. 5 (b), as

$$\begin{aligned} x'(m) &= \int_0^T s(t)\psi_m^*(t)\mathrm{d}t \\ &= \sum_{k=0}^{N-1} x(k)\delta(m-k) = x(m). \end{aligned} \tag{18}$$

## C. Digital Implementation of OCDM

The OCDM system can be implemented digitally, as shown in Fig. 6. The discrete time-domain OCDM signal is obtained by sampling the continuous time-domain signal in Equ. (17). Since there are two forms of DFnT matrix in Equ. (6) depending on $N$ being even or odd, the discrete OCDM signal is, if $N$ is even,

$$\begin{aligned} s(n) &= s(t)\big|_{t=n\frac{T}{N}} = \sum_{k=0}^{N-1} x(k)\psi_k\left(n\frac{T}{N}\right) \\ &= e^{j\frac{\pi}{4}}\sum_{k=0}^{N-1} x(k)e^{-j\frac{\pi}{N}(n-k)^2}, \end{aligned} \tag{19}$$

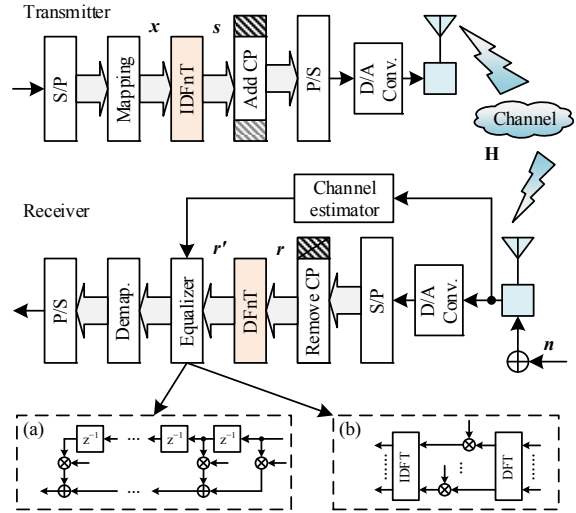

Fig. 6. Schematic diagram of the proposed digital implementation of OCDM. Insets: Illustrations of (a) time-domain and (b) frequency domain equalizer.

and, if $N$ is odd,

$$s(n) = s(t)\big|_{t=\left(n+\frac{1}{2}\right)\frac{T}{N}} = \sum_{k=0}^{N-1} x(k)\psi_k\left(n\frac{T}{N}\right)$$
$$= e^{j\frac{\pi}{4}}\sum_{k=0}^{N-1} x(k)e^{-j\frac{\pi}{N}\left(n-k+\frac{1}{2}\right)^2}. \tag{20}$$

Inspecting Equ. (19) and (20), one can compare them with the definition of DFnT in Equ. (6), and find that they are exactly the IDFnT. Thus, the synthesis of a bank of discretized modulated chirp waveforms can be realized by the IDFnT. To express Equ. (19) and (20) in a concise matrix form, we stack the symbols in the vector form as $\boldsymbol{x} = [x(0), x(1), \ldots, x(N-1)]^T$, and thus the discrete time-domain OCDM signal is

$$\boldsymbol{s} = \boldsymbol{\Phi}^H \boldsymbol{x}. \tag{21}$$

Since the DFnT matrix is unitary, at the receiver the transmitted information symbols can be recovered by performing the inverse operation, i.e., DFnT. The recovered symbols are thus

$$\boldsymbol{x}' = \boldsymbol{\Phi}\boldsymbol{s} = \boldsymbol{x}. \tag{22}$$

## IV. OCDM Signal Under LTI Systems

In this section, we will formulate the mathematical model of the OCDM signal under the LTI

channel with additive white Gaussian noise (AWGN) in matrix form. The channel is static or quasi-static, which means that the channel response remains constant within one OCDM block, and might change in the next block. It is assumed that the channel information is available at the receiver via some channel estimation method with perfect timing and frequency synchronization. To avoid inter-symbol interference (ISI), guard interval is inserted between adjacent blocks, just as the OFDM system does. Based on the analysis, we will show that the guard interval can be filled with either zeros, i.e., zero-padded prefix (ZP), or the replica of a portion of the end of the signal, i.e., cyclic prefix (CP).

### A. Signal Transmission under LTI Gaussian Channels

In Fig. 6, the signal transmission is illustrated. Suppose that the maximum channel delay is smaller than the length of guard interval. If CP is used for guard interval, the signal experienced the channel $\mathbf{H}$ with AWGN is

$$\mathbf{r} = \mathbf{H}\mathbf{s} + \mathbf{n} = \mathbf{H}\mathbf{\Phi}^H \mathbf{x} + \mathbf{n} \qquad (23)$$

where $\mathbf{H}$ is the channel impulse response (CIR) matrix and $\mathbf{n}$ is the AWGN vector. The CIR matrix $\mathbf{H}$ is circulant, and its first column is $\mathbf{h} = [h(0), h(1), \ldots, h(L-1), 0, \ldots, 0]^T$, where $h(l), l = 0, \ldots L-1$, are the CIR taps and $L$ is the maximum delay spread.

On the other hand, if ZP is used, one can also arrive at Equ. (23) by overlap-and-add operation, [Chapter 12, 51]. Therefore, Equ. (23) can be a general model for the OCDM signal transmitting under LTI channels based on both CP and ZP.

Before we recover the symbols, we could first compensate the channel $\mathbf{H}$ in Equ. (23), and then perform DFnT to recover the transmitted symbols in $\mathbf{x}$. Alternatively, we first perform DFnT on the received signal as

$$\mathbf{r}' = \mathbf{\Phi}\mathbf{r} = \mathbf{\Phi}\mathbf{H}\mathbf{s} + \mathbf{\Phi}\mathbf{n}. \qquad (24)$$

Based on and the convolution property of DFnT in Equ. (8), Equ. (24) is further given by

$$r' = \Phi H \Phi^H x + \Phi n$$
$$= Hx + \Phi n. \tag{25}$$

Therefore, the chirp waveforms $\Phi^H$ vanish and are transparent to the channel $H$ as if symbols are transmitted directly without modulating the chirps after the DFnT at the receiver. In addition, since $\Phi$ is unitary, the noise $\Phi n$ is still AWGN.

From Equ. (25), the CIR $H$ can be compensated by using the multi-tap time-domain equalization (TDE) in Fig. 6 (a) or by using the more efficient single-tap FDE in Fig. 6 (b) to recover the symbols $x$. More sophisticated nonlinear equalizers or decoding algorithms, e.g., decision feedback equalizer (DFE) and maximal likelihood (ML) detection, are more powerful for the improvement of performance [52, 53]. However, they complicate the computational and hardware complexity, and we only consider the linear equalizers in this paper.

### B. Proposed Equalization Algorithm for OCDM

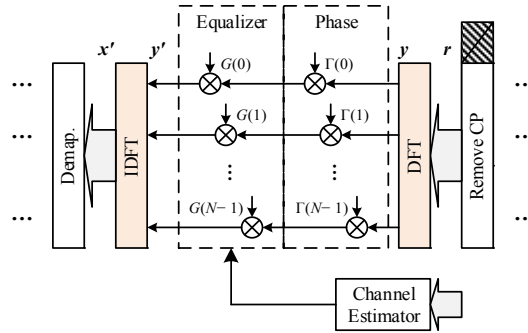

Fig. 7. Proposed single-tap FDE for the OCDM system.

In this subsection, we propose an efficient algorithm based on the FDE for compensating the channel dispersion imposed on the OCDM signal. The diagram of the proposed algorithm is illustrated in Fig. 7. In the proposed scheme, single-tap equalization is employed, and the IDFnT at the receiver is avoided based on its eigen-decomposition property.

At the receiver, the sampled signal is given in Equ. (23). Instead of DFnT in Fig. 6, it is first transformed into the frequency domain by DFT as

$$\boldsymbol{y} = \mathbf{F}\boldsymbol{r} = \mathbf{FH}\boldsymbol{\Phi}^H \boldsymbol{x} + \mathbf{F}\boldsymbol{n}, \tag{26}$$

where $\mathbf{F}$ is the normalized DFT matrix. Using the identity $\mathbf{I} = \mathbf{F}^H\mathbf{F}$, Equ. (26) can be further given by

$$\begin{aligned} \boldsymbol{y} &= \mathbf{FHF}^H \mathbf{F}\boldsymbol{\Phi}^H \mathbf{F}^H \mathbf{F}\boldsymbol{x} + \boldsymbol{w} \\ &= \boldsymbol{\Lambda}\boldsymbol{\Gamma}^H \mathbf{F}\boldsymbol{x} + \boldsymbol{w}, \end{aligned} \tag{27}$$

where $\boldsymbol{\Lambda} = \mathbf{FHF}^H$ is the channel frequency response (CFR) matrix and $\boldsymbol{\Gamma}^H = \mathbf{F}\boldsymbol{\Phi}^H\mathbf{F}^H$ is a coefficient matrix.

According to the eigen-decomposition property of a circulant matrix with respect to DFT, the CFR matrix $\boldsymbol{\Lambda}$ is diagonal. The $k$-th diagonal entry of $\boldsymbol{\Lambda}$ is the CFR at the $k$-th frequency bin, or equivalently is the $k$-th eigenvalue to the $k$-th column vector of $\mathbf{F}^H$.

Recalling Equ. (6), the DFnT matrix is also circulant. Thus, the matrix $\boldsymbol{\Gamma}$ is also a diagonal matrix whose diagonal entries are the eigenvalues of $\boldsymbol{\Phi}$ with respect to $\mathbf{F}^H$. The eigenvalue of $\boldsymbol{\Phi}$ is given in [46], as

$$\Gamma(k) = \begin{cases} e^{-j\frac{\pi}{N}k^2} & N \equiv 0 (\mathrm{mod}\,2) \\ e^{-j\frac{\pi}{N}k(k-1)} & N \equiv 1 (\mathrm{mod}\,2) \end{cases}. \tag{28}$$

Based on the commutative law of the product of two diagonal matrices, Equ. (27) can be further given by

$$\boldsymbol{y} = \boldsymbol{\Gamma}^H \boldsymbol{\Lambda} \mathbf{F}\boldsymbol{x} + \boldsymbol{w}. \tag{29}$$

Before compensating the CFR $\boldsymbol{\Lambda}$, we first cancel out the phase induced by $\boldsymbol{\Gamma}$. The equalized signal is given by

$$\boldsymbol{y}' = \mathbf{G}\boldsymbol{\Gamma}\boldsymbol{y} = \mathbf{G}\boldsymbol{\Lambda}\mathbf{F}\boldsymbol{x} + \mathbf{G}\boldsymbol{\Gamma}\boldsymbol{w}, \tag{30}$$

where $\mathbf{G}$ is a diagonal matrix with its $k$-th diagonal entry $G(k)$ to be the coefficients of the single-tap equalizer. For example, if zero-forcing (ZF) criterion is adopted, $G(k)$ is

$$G_{ZF}(k) = \Lambda^{-1}(k), \tag{31}$$

and if minimum mean square error (MMSE) is adopted, $G(k)$ is

$$G_{MMSE}(k) = \frac{\Lambda^*(k)}{\left|\Lambda(k)\right|^2 + \rho^{-1}}, \qquad (32)$$

where $\rho$ is the signal-to-noise ratio (SNR). Finally, the signal is transformed by IDFT to recover the transmitted information. If the ZF equalizer is employed, the recovered signal is

$$\boldsymbol{x}'_{\text{ZF}} = \boldsymbol{x} + \mathbf{F}^H \boldsymbol{\Lambda}^{-1} \boldsymbol{\Gamma} \boldsymbol{w}, \qquad (33)$$

and, on the other hand if MMSE is used, the signal is

$$\boldsymbol{x}'_{\text{MMSE}} = \left( \mathbf{F}^H \frac{\boldsymbol{\Lambda}^H \boldsymbol{\Lambda}}{\boldsymbol{\Lambda}^H \boldsymbol{\Lambda} + \rho^{-1} \mathbf{I}} \mathbf{F} \right) \boldsymbol{x}$$
$$+ \mathbf{F}^H \frac{\boldsymbol{\Lambda}^H}{\boldsymbol{\Lambda}^H \boldsymbol{\Lambda} + \rho^{-1} \mathbf{I}} \boldsymbol{\Gamma} \boldsymbol{w}. \qquad (34)$$

Although the ZF equalizer is able to completely remove the channel distortion, it is notorious for noise enhancement. In practice, the MMSE equalizer can efficiently balance the noise enhancement and channel compensation.

## V. DISCUSSIONS

From previous sections, we can see that Fresnel transform mathematically formulates the OCDM, just as the Fourier transform in OFDM. Inspecting Equ. (1) and (6), the Fresnel transform and DFnT are trigonometric transforms with quadratic phases. Both the Fresnel transform and Fourier transform are LCT, and they are intimate to each other [31-34].

In this section, we will study the algebraic properties of the Fresnel transform and the Fourier transform. According to the properties, the implementation differences and compatibility of the OCDM and the OFDM are discussed. It is shown that the DFnT can be realized by the DFT using fast Fourier transform algorithms, and thus the OCDM system can be realized by the existing OFDM system without complicating the complexity.

*A. Relation between the Fourier and Fresnel transforms*

In the continuous case, the kernel of the most 'well-known' Fourier transform is

$$\omega(f, t) = e^{-j2\pi ft}, \tag{35}$$

and the expanded kernel of Fresnel transform in Equ. (1) is

$$\varphi_a(f, t) = e^{-j\frac{\pi}{4}} e^{j\frac{\pi}{a}\left(f^2 - 2ft + t^2\right)}. \tag{36}$$

In the discrete case, they are

$$W(m, n) = \frac{1}{\sqrt{N}} e^{j\frac{2\pi}{N}mn}, \tag{37}$$

and

$$\Phi(m, n) = \frac{1}{\sqrt{N}} e^{-j\frac{\pi}{4}}$$
$$\times \begin{cases} e^{j\frac{\pi}{N}\left(m^2 - 2mn + n^2\right)} & N \equiv 0 \,(\mathrm{mod}\, 2) \\ e^{j\frac{\pi}{4N}} e^{j\frac{\pi}{N}\left(m^2 + m - 2mn + n^2 - n\right)} & N \equiv 1 \,(\mathrm{mod}\, 2) \end{cases}, \tag{38}$$

respectively. The kernel of Fresnel transform in Equ. (36) or DFnT in Equ. (38) contains the kernel of Fourier transform or DFT with the additional quadratic phases. In the discrete form, the additional quadratic phases are

$$\Theta_1(m) = e^{-j\frac{\pi}{4}} \times \begin{cases} e^{j\frac{\pi}{N}m^2} & N \equiv 0 \,(\mathrm{mod}\, 2) \\ e^{j\frac{\pi}{4N}} e^{j\frac{\pi}{N}\left(m^2 + m\right)} & N \equiv 1 \,(\mathrm{mod}\, 2) \end{cases}, \tag{39}$$

and

$$\Theta_2(n) = \begin{cases} e^{j\frac{\pi}{N}n^2} & N \equiv 0 \,(\mathrm{mod}\, 2) \\ e^{j\frac{\pi}{N}\left(n^2 - n\right)} & N \equiv 1 \,(\mathrm{mod}\, 2) \end{cases}, \tag{40}$$

Consequently, the DFnT can be implemented by FFT in three steps:

1) multiplying the chirp phase $\boldsymbol{\Theta}_2$,

2) performing the DFT by FFT algorithm, and finally

3) multiplying the other chirp phase $\mathbf{\Theta}_1$,

where $\mathbf{\Theta}_1$ and $\mathbf{\Theta}_2$ are diagonal matrices whose $m$-th diagonal entries are $\Theta_1(m)$ and $\Theta_2(m)$, respectively.

## B. Compatibility of the OCDM to OFDM

In the previous subsection, it was shown that the Fresnel transform or DFnT is divided into a three-step process involving Fourier transform or DFT. So one would expect that the OCDM scheme can be integrated into the OFDM system easily without significant modification on the OFDM transceiver.

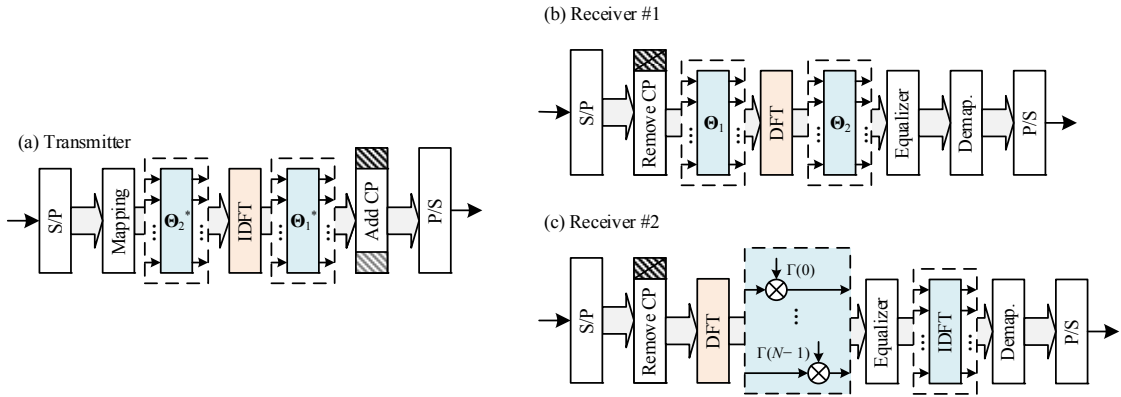

Fig. 8. Schematic diagram of OFDM system (excluding the dash components) and the diagram of OCDM system based on the OFDM system (including the dash components).

Fig. 8 provides the system diagram of a conventional OFDM system (without the components in the dashed-line boxes). At the transmitter, the IDFT multiplexes symbols onto the parallel sub-channels, and at the receiver DFT performs inverse operation to recover the symbols. After the DFT at the receiver side, single-tap equalizer compensates each subchannel.

According to the relation between the DFT and the DFnT in Equ. (35)-(40), the OCDM can be integrated into the OFDM system with the additional operations in the dashed-line boxes. At the transmitter, the three-step operation involving IDFT acts as the IDFnT. At the receiver end, there are two architectures. One is based on the receiver in Fig. 6, and the other is based on the proposed equalization algorithm in Section IV, refer to Fig. 7. In the receiver #1, the three-step operation functions as DFnT. It should be noted that, as discussed in Section III-C, the equalizer for such scheme could be either multi-tap TDE in Fig. 6 (a) or FDE in Fig. 6 (b). If the FDE is adopted, in addition to the single-tap equalizer, there is a DFT and an IDFT operation. The receiver #2 is the same as that in Fig. 7, and only single-tap equalizer is required.

In terms of the signal structure, both the OFDM and OCDM systems transmit the synthesized modulated waveforms in blocks. Between the blocks, guard interval is used to avoid ISI. As shown in Section III-C, both CP and ZP can be used for filling the guard interval in the OCDM system. The structure of OCDM signal is the same as that of the OFDM signal. Therefore, the structure of the OCDM signal is compatible to the OFDM signal. The design of OCDM system can be well integrated into the OFDM system.

In summary, the generation and recovery of OCDM signal can be realized using the existing OFDM system with additional operations involving only phase rotation and one more IDFT at the receiver. In the following subsection, we will compare the arithmetical complexity between the OCDM and OFDM systems.

## C. Arithmetic Complexity of OCDM

In Fig. 8, the similarities and differences between the OFDM and OCDM systems are illustrated. In the OCDM system, there are two receiver schemes. For both schemes, the transmitter is the same, and thus the complexity of OCDM system depends on which receiver scheme is adopted. In this subsection, we compared the additional arithmetic complexity of the OCDM system to the OFDM system in terms of complex multiplication operations. It should be noted that in a communication system, there are other compulsory modules and operations, such as synchronizations and channel estimation, which require additional consideration. The arithmetic complexities of those vary depending on the algorithms adopted. In this paper, we will not provide these details for brevity.

At the transmitter, there are two additional phase rotations, which takes additional 2 complex multiplications per symbol compared to the transmitter of OFDM system.

If the receiver #1 is adopted, there are also two phase rotation operations. In the equalization module, either TDE or FDE can be adopted, Fig. 6 (a) and (b). The complexity of TDE depends on the number of taps of transverse filter. If the number of taps is $L$, which is larger than the CIR taps, the complexity of the TDE is $L$ per symbol. On the other hand, in the FDE scheme, besides the single-tap equalizer, it needs two more DFT operations, see Fig. 6 (b). Therefore, the FDE scheme requires additional $\log_2 N$ complex multiplications per symbol. In the applications, such as the radio mobile communication, whose channel delay spread is relatively large, the FDE is more preferable than the TDE in terms of the computation complexity.

In the receiver scheme #2, besides the phase cancellation and the single-tap equalizer, an additional IDFT are required. Thus, the additional complexity compared to the OFDM system is $0.5\log N$.

TABLE I
ADDITIONAL ARITHMETIC COMPLEXITY OF THE OCDM SYSTEM
COMPARED TO THE OFDM SYSTEM

| Transmitter | Receiver #1 | | Receiver #2 |
|---|---|---|---|
| | TDE | FDE | |
| 2 | $L + 2$ | $\log_2 N + 2$ | $\frac{1}{2} \log_2 N$ |

The arithmetical complexity is evaluated by complex multiplications per subcarrier/chirp. $L$ is the number of taps of the time-domain equalizer (TDE), and $N$ is the number of chirps/subcarriers in OFDM/OCDM system.

The additional arithmetical complexity of the OCDM system compared to the OFDM system is provided in TABLE I. It can be seen that the complexity of the OCDM system is slightly increased to that of the OFDM if the receiver #2 is adopted.

VI. Simulation

To investigate the feasibility and performance of the OCDM system, simulations are performed under wireless channel. The system bandwidth is 10 MHz, and there are 1024 chirps modulated in $M$-ary QAM with $M$ from 4 to up to 64. In the OCDM system, the guard interval is filled with CP, and its length is chosen larger than the maximum excess delay of the channel to avoid ISI. Both the ZF and MMSE equalizers are adopted to show the effect of noise enlargement in the OCDM system with linear equalizers. In the simulation, OFDM is investigated for comparison. Similarly, the OFDM system has 10-MHz bandwidth which is divided into 1024 subcarriers modulated in QAM. CP is also used for OFDM to fill the guard interval. In the OFDM, both the ZF and MMSE equalizers achieves the same bit-error rate (BER) performance because symbols are communicated in parallel subchannels. Thus, the ZF equalizer is employed for OFDM system for simplicity.

TABLE II
POWER DELAY PROFILE OF THE EVA MODEL

| Excess tap delay (ns) | 0 | 30 | 150 | 310 | 370 | 710 | 1090 | 1730 | 2510 |
|---|---|---|---|---|---|---|---|---|---|
| Relative Power (dB) | 0 | −1.5 | −1.4 | −3.6 | −0.6 | −9.1 | −7.0 | −12.0 | −16.9 |

In the simulation, two channel models are applied: 1) the multipath Rayleigh fading channel model with 10-ray and 2) the more practical extended vehicular A (EVA) channel model. In the 10-ray multipath Rayleigh fading channel, the paths have equal gain and uniform distributed delay. The maximum excess delay spread is 5.4 μs. In TABLE II, the power delay profile of EPA model is provided. Both the channels are quasi-static; it means that channel response remains static within one OCDM/OFDM block and it varies for the next. The thermal noise modeled as AWGN is measured at the receiver. The BER performance is evaluated via signal-to-noise ratio (SNR). In this paper, the SNR is given by the energy per bit $E_b$ to the noise power density $N_0$, ($E_b / N_0$).

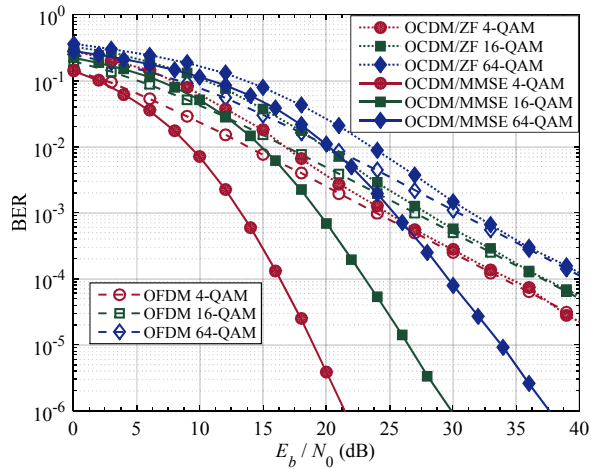

Fig. 9. BER performance of the OCDM systems with both ZF and MMSE equalizers and the OFDM system under the 10-ray multipath Rayleigh fading channel.

In Fig. 9, the BER performance is provided in the 10-ray multipath Rayleigh fading channel. For the OCDM system with ZF equalizer, higher SNR is required to achieve the same BER as the OFDM system does, especially when the SNR is small. The corresponding BER curves approach those of the OFDM as the SNR increases. This degradation is caused by the noise enhancement of the ZF equalizer, which becomes smaller as SNR increases.

The OCDM system with MMSE equalizer outperforms the OFDM since MMSE equalizer is able to balance the channel compensation and noise enhancement in the OCDM system. The multipath diversity in the OCDM contributes to the superior performance over OFDM under multipath fading channel. It should be noted that the performance of the OCDM system with MMSE is inferior to the OFDM system in the low SNR region, and the degradation is more pronounced as the modulation level increases from 4- to 64-QAM. In the OCDM system with MMSE equalizer, the noise enhancement still exists, and they are more severe in the low SNR region. In addition, high-level modulation is more sensitive to the detrimental effects. These account for the degradation of the OCDM system with MMSE equalizer compared to the OFDM system in the low SNR region. Nonetheless, the OCDM system with MMSE equalizer still gets much better performance in higher SNR for its ability of utilizing the multipath diversity.

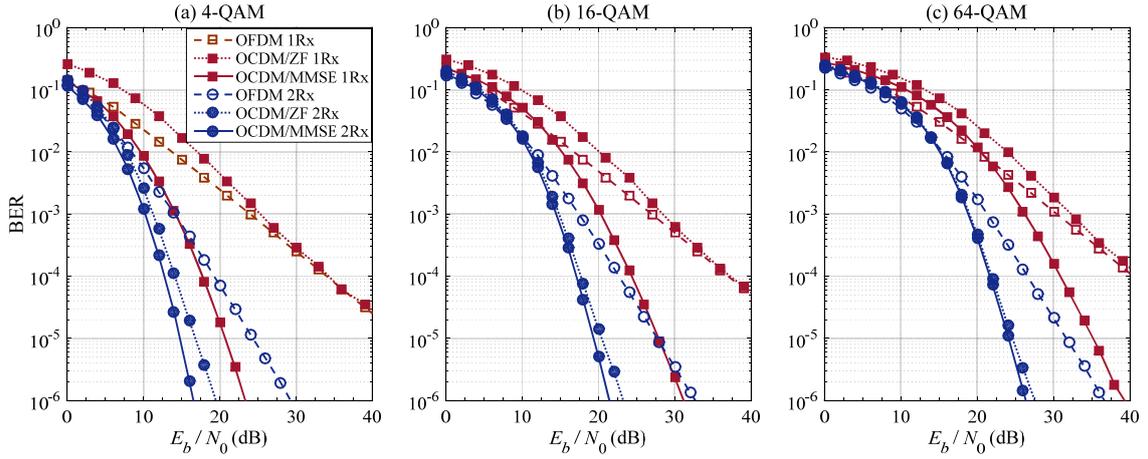

Fig. 10. BER performance of the OCDM systems with both ZF and MMSE equalizers and the OFDM system under the LTE extended vehicle A channel model with receiver diversity; (a) 4-QAM, (b) 16-QAM, and (c) 64-QAM.

The BER performance under the EVA channel is provided in Fig. 10. In addition, receive diversity with two antennas based on maximum ratio combining (MRC) is adopted. The received power is normalized rather than doubled by two antennas. By doing so, the SNR of the received signal after MRC remains unchanged for both single and two receive antennas to show the effect of spatial diversity in the OCDM system. If the receive power is not normalized, 3-dB SNR improvement will be observed over the normalized case.

Similar results to Fig. 9 can be observed in the EVA channel without receive diversity. However, if receive diversity applies, there is significant performance improvement for both OCDM and OFDM systems. The improvement is more notable for the OCDM system of ZF equalizer. With 2 Rx, the performance of OCDM with ZF is much better than that of the OFDM system, and it even approaches the MMSE case. For the OCDM systems of ZF and MMSE, the effect of noise enhancement is efficiently suppressed by spatial diversity, and their performance in the low SNR region is also improved compared to the single antenna case. For example, in the low SNR region, the performance of the OCDM system is inferior to that of OFDM system without spatial diversity, while with spatial diversity the performance of OCDM is similar to that of OFDM. The results imply that spatial diversity can significantly suppress the effect of noise enhancement in the OCDM system with linear equalizers, and the multipath diversity dominates the performance.

## VII. Conclusion

In this paper, we presented the principle of orthogonally multiplexing a bank of chirp waveforms whose amplitude and phase are used for modulation. The principle is based on the Fresnel transform, and the convolution theorem of the Fresnel transform gives the analytical approach to model the transmission of OCDM signal in LTI channels. The digital implementation of OCDM system is proposed based on DFnT, and it is shown that the chirp waveforms are transparent to the dispersive channel. Both the time-domain and frequency-domain equalizers can be applied for channel compensation. By exploiting the eigen-decomposition of the DFnT matrices, a simpler and more efficient single-tap equalization algorithm is proposed for the OCDM system. In terms of the implementation of OCDM system, it is shown that the system is compatible to the OFDM system, and it can be realized using the existing OFDM system without significant modification.

Simulations under wireless multipath channel are carried out to validate the feasibility of the OCDM. The results show that the OCDM is able to exploit the multipath diversity with linear equalizers and the OCDM with MMSE equalizer outperforms the OFDM system. Furthermore, with the spatial diversity, the noise enhancement of linear equalizers can be efficiently suppressed in the OCDM system. By exploiting the receive diversity, even the OCDM system with ZF equalizer achieves notable performance improvement compared to the OFDM system.

Consequently, for its compatibility to the widespread OFDM system and its capability to counteract the detrimental effects in communication channels, the OCDM system is an attractive alternative solution for the high-speed communication systems.